\documentclass[aps,prl,preprint,superscriptaddress]{revtex4-2}

\usepackage{amsmath}
\usepackage{slashed}

\begin{document}


\title{No Future in Black Holes}

\author{Malcolm J. Perry}
\email{malcolm@damtp.cam.ac.uk}
\affiliation{School of Physics and Astronomy, Queen Mary University of London, Mile End Road, London E1 4NS, UK}
\affiliation{Centre for Mathematical Sciences, Wilberforce Road, Cambridge, CB3 0WA, UK}
\affiliation{Trinity College, Cambridge, CB2 1TQ, UK.}

\date{\today}

\begin{abstract}
The black hole information paradox has been with us for some time. We outline the nature of the paradox. 
We then propose a resolution based on an examination of the properties of quantum gravity
under circumstances that give rise to a classical singularity. We show that the gravitational wavefunction
vanishes as one gets close to the classical singularity. This results in a future boundary condition 
inside the black hole that allows for quantum information to be recovered in the evaporation process.
\end{abstract}


\maketitle


The black hole information paradox is presently one of the most puzzling problems in fundamental physics.
It is widely believed that quantum mechanics should control the evolution of any physical system. 
At first sight, black holes appear to transcend quantum mechanics, \cite{Hawking:1976ra}. 

What follows is
a brief outline of a possible way to resolve the information paradox. A much more complete 
treatment will be published elsewhere.

In classical physics, a black 
hole formed by gravitational collapse will settle down to a member of the  Kerr-Newman family. 
The black hole is completely characterised by its mass, $M$, angular momentum $J$,
electric charge $Q$ and its soft gauge charges.\footnote{We use natural units with $G=c=\hbar=k=1$.
The curvature convention is $\nabla_a\nabla_b V_c = R_{abcd}V^d$.}
Infalling matter, 
once it has passed through the event horizon can
no longer influence anything outside the black hole. What is more, it will inevitably arrive at the 
spacetime singularity in the black hole interior and disappear from the spacetime. From the point of view of
external observers, the material that gave rise to the black hole has completely disappeared. The black
hole will then remain as a sink for infalling material for ever. 

In quantum mechanics, the situation is different because of Hawking radiation, 
\cite{Hawking:1974sw,Hawking:1974rv}. Hawking showed that black 
holes produce thermal radiation at a temperature inversely proportional to their mass. A black hole
therefore loses mass more and more rapidly and will disappear completely on a timescale $\tau \sim M^3$.
The difficulty is that the final state is just thermal
Hawking radiation that appears to have a von Neumann entropy $\sim M^2$ and is independent of the nature 
of the incoming state. In the unitary evolution
required by quantum mechanics, von Neumann entropy is constant. So either quantum mechanics 
breaks down or some new physics is required.

The remainder of this paper presents a concrete proposal, based on the quantization of general
relativity, that allows for a possible resolution of the paradox with the black hole obeying the usual rules of 
quantum mechanics as far as external observers are concerned.  We will restrict the discussion to 
a spacetime of dimension four and only consider black holes in
an asymptotically flat spacetime. 

The hoop conjecture \cite{Thorne:1972ji} asserts that a black hole will form when a mass $m$ is 
localised into a sufficiently 
small region of size $\sim m$. Once an event horizon forms, Penrose's theorem \cite{Penrose:1964wq}
guarantees that 
the spacetime is singular and is usually taken to mean that there is a region of infinite
curvature that is the boundary of spacetime. 

The special case of spherically symmetric collapse was studied by Oppenheimer and Snyder, 
\cite{Oppenheimer:1939ue}. They looked at
the collapse of a spherically symmetric ball of pressure-free matter.
Outside the collapsing body, the metric is given precisely by the static Schwarzschild metric. The
horizon forms once the ball has contracted sufficiently. The spacetime inside the ball of matter is 
part of a $k=1$ FLRW universe. The singularity forms to the future of the horizon when the
density of the ball diverges. 
However, the singularity is not quite the same as in the familiar FLRW universe as it stretches outside the
collapsing matter in a spacelike fashion. 

Realistic gravitational collapse is not spherically symmetric. Weak cosmic censorship is the conjecture
that any singularity is hidden from observers outside the black hole. Whilst a proof of this conjecture
is lacking, the evidence for it is substantial.  Unlike spherical collapse
to form a Schwarzschild black hole, the Kerr-Newman solutions involve timelike singularities.
It is believed that they do not form in realistic collapse. They are behind an inner horizon 
which is known to be perturbatively unstable. Dafermos and Luk \cite{Dafermos:2017dbw}
have recently shown that generically
singularities have spacelike, or null, components. We assume the Penrose diagram for generic collapse
to be the similar to the Schwarzschild case although there are possibly null sections
of the singularity. The essential point being that we expect all future-directed timelike or null lines 
to reach the singularity. 

Hawking \cite{Hawking:1974sw, Hawking:1974rv} showed that the temperature of the black hole is given 
by $T_H={\kappa/2\pi}$ where $\kappa$
is the surface gravity of the black hole. Bardeen, Carter and Hawking \cite{Bardeen:1973gs} 
proved the first law of black 
hole mechanics that governs infinitesimal changes in the state of a stationary black hole.
\begin{equation} dM=\frac{\kappa dA}{8\pi} + \Phi dQ + \Omega dJ \end{equation}
where $M$ is the mass of the black hole, $A$ the area of the event horizon, $\Phi$ 
the electrostatic potential of the hole and
$\Omega$ its angular velocity. Since $T_H={\kappa/2\pi}$, we readily infer 
the black hole  has entropy ${A/4}$. It should be noted that $S\sim M^2$ is an entropy that is vastly 
greater than typical systems in equilibrium for which the entropy scales with mass more slowly.

Following Boltzmann, it is presumed that  $e^S$ is the density of states of the black hole.
This last idea is formalised in a collection of ideas that have become known as 
the \lq\lq central dogma\rq\rq\ of black
hole physics \cite{Almheiri:2020cfm}.
For external observers, the black hole behaves like any other quantum system;
it has a density of states $e^S$ and its evolution is unitary. Outside the black hole, one expects
the conventional ideas of general relativity to be valid; namely that spacetime can be treated by
classical geometry and that fields (including gravitons) can be treated by effective field theory.

If these ideas were to hold inside a black hole we would be in trouble. Suppose we study the wave equation 
inside the Schwarzschild black hole, we find that its solutions typically blow up at the singularity
and induce a corresponding divergence in the probability current there. Quantum information 
is thereby taken out of the spacetime. In the absence of cloning, 
information would be lost.

It is to quantum gravity that we must turn our attention. The path integral for gravity is
\begin{equation} Z \sim \int {\mathcal D}g \ e^{iI[g]} \end{equation}
where $I[g]$ is the Einstein action for a metric $g_{ab}$ on a manifold ${\mathcal M}$ and boundary
$\partial{\mathcal M}$ with induced metric $\gamma_{ij}$,
\begin{equation} I[g] = \frac{1}{16\pi}\int_{\mathcal M} R\ \sqrt{-g}\ d^4x  \pm \frac{1}{8\pi}
\int_{\partial{\mathcal M}}
K\ \sqrt{\pm \gamma}\ d^3x. \end{equation} 
$R$ is the Ricci scalar of the metric $g_{ab}$. $K$ is the trace of the second fundamental form of
the boundary and the sign is chosen depending on whether the boundary is spacelike or timelike. 
There may be many disconnected components of the boundary. The integral is taken over all 
Lorentzian metrics
$g_{ab}$ modulo diffeomorphisms with induced metric $\gamma_{ij}$ on the boundary. 

It is not clear what, if any, precise conditions are to be imposed on the metric. If we are
interested only in a semi-classical evaluation of $Z$, the path integral can be approximated using
Picard-Lefshetz theory \cite{Witten:2010cx} 
in which case it seems to be necessary to deform $g_{ab}$ into the complex.
The use of the path integral is limited by the failure of renormalizability of the Einstein theory.
The physical meaning of $Z$ is that is gives the probability amplitude of finding a configuration with
metric $\gamma_{ij}$ on the boundary, or boundaries, of ${\mathcal M}$. As an example, $Z$ might be the
transition amplitude for having a metric $\gamma^{(1)}$ on one surface and $\gamma^{(2)}$ on another,
\cite{Hawking:1980gf}. In quantum gravity, when  describing a closed system, there is no way 
of determining whether $\gamma^{(1)}$ is to the past or future of $\gamma^{(2)}$. If however the 
surfaces on which $\gamma^{(i)}$
are defined stretch out to infinity, then a time can be associated with each such surface. 
A second possibility is that there is a single component to the boundary. $Z$ is then the probability 
amplitude of finding a particular geometry $\gamma$, \cite{Hartle:1983ai}. In the case of a single 
boundary component, $Z$ is usually referred
to as the wavefunction of the universe, $\Psi[\gamma]$. Again, there is no 
reference to an arrow of time for a closed system. Inside an evaporating black hole, if one has a 
surface close to the singularity, 
it does not seem to be legitimate to extend such a surface out to infinity as the interior is causally 
disjoint from the asymptotic region. 

Canonical methods allow for another approach to determining $\Psi[\gamma]$. Take the metric and rewrite
it in ADM form \cite{Arnowitt:1962hi,Dirac:1958sc} as
\begin{equation} ds^2 = -N^2dt^2 + \gamma_{ij}(dx^i+N^idt)(dx^j+N^jdt). \end{equation}
$N$ is referred to as the lapse and $N^i$ as the shift. $\gamma_{ij}$ is a purely spacelike metric. 
Using this decomposition, the Einstein action becomes
\begin{equation} I[g] = \int d^3x\,dt\ \sqrt{\gamma}\, N\, \Bigl(K_{ij}K^{ij} - K^2 +
 {}^{(3)}R(\gamma)\Bigr) \end{equation}
where  ${}^{(3)}R(\gamma)$ is the Ricci scalar of the three-metric $\gamma$ and $K_{ij}$
is the second fundamental form of the $t=const$ surfaces. Explicitly
\begin{equation} K_{ij} = \frac{1}{2N}\bigl(D_iN_j +  D_jN_i - \dot\gamma_{ij}\bigr) \end{equation}
where $D_i$ is the covariant derivative based on the $3$-metric $\gamma_{ij}$
and a dot denotes the time derivative. Using Hamiltonian
techniques, one finds that in the Gaussian gauge $N=1, N^i=0$ the system is described entirely in terms
of two constraints: the diffeomorphism constraint
\begin{equation} \chi^j \equiv D_i\pi^{ij} =0 \end{equation} 
and the Hamiltonian constraint
\begin{equation} {\mathcal H} \equiv \gamma^{-{1/2}}\bigl(\pi^{ij}\pi_{ij}-\tfrac{1}{2}\pi^i_i\pi^j_j
-{}^{(3)}R\gamma\bigr) = 0. \end{equation}
$\pi^{ij}$ is the momentum conjugate to $\gamma_{ij}$ and can be written as
\begin{equation} \pi^{ij} = -\gamma^{1/2}(K^{ij}-\gamma^{ij}K). \end{equation}
One quantizes the system by replacing $\pi^{ij}$ by $-i{\delta/{\delta\gamma_{ij}}}$. The Hamiltonian 
constraint then becomes the Wheeler-DeWitt equation \cite{Wheeler:xx,DeWitt:1967yk}
\begin{equation} \Bigl(G_{ijkl}\frac{\delta}{\delta\gamma_{ij}}\frac{\delta}{\delta\gamma_{kl}} -
\gamma^{1/2} {}^{(3)}R\Bigr)\ \Psi[\gamma] = 0.
\end{equation}
$G_{ijkl}$ is the DeWitt co-metric
\begin{equation} G_{ijkl} =\tfrac{1}{2}\gamma^{-{1/2}}\bigl(\gamma_{ik}\gamma_{jl}+\gamma_{il}\gamma_{jk}
-\gamma_{ij}\gamma_{kl}\bigr).\end{equation}
$\Psi[\gamma]$ here is the same object as that defined by the path integral and is a functional
on superspace,
the space of all spatial metrics modulo diffeomorphisms. 
\footnote{The equivalence 
however is rather formal as there are ambiguities in the path integral measure, operator ordering 
and issues over what domain the fields are defined.}

When one studies systems that are far from equilibrium, it is  convenient to suppose the initial
state is specified by a density matrix $\rho^{(i)}$ rather than a pure state with some specific geometry
$\gamma$.
The usual path integral is then generalised by a method of Schwinger \cite{Schwinger:1960qe}, 
Keldysh \cite{Keldysh:1964ud}, and  Kadanoff and Baym \cite{Kadanoff:xx}. The path integral defines 
the probability $P[\gamma^{(f)},\rho_i]$
of finding a final state geometry
$\gamma^{(f)}$ given the initial density matrix $\rho^{(i)}$.
\begin{equation} P[\gamma^{(f)},\rho_i] = 
{\mathcal N}\ {\rm Tr} \int {\mathcal D}g^{(+)}\ {\mathcal D}g^{(-)}
\ e^{iI[g^{(+)}]}\rho^{(i)}e^{-iI[g^{(-)}]}. \end{equation}
$\rho^{(i)}=\rho^{(i)}_{\gamma\gamma^\prime}\vert\gamma\rangle\langle\gamma^\prime\vert$
where $\gamma$ and $\gamma^\prime$ refer to some geometries.
The forward part of the path integral is over spacetime metrics $g^{(+)}$ with boundary $\gamma^{(f)}$ 
in the future and $\gamma$ in the past. The time-reversed path integral is over $g^{(-)}$
with $\gamma^{(f)}$ in the future and $\gamma^\prime$ in the past.  ${\mathcal N}$
is a normalization constant. 

Aharonov, Bergmann and Lebowitz \cite{Aharonov:xx} observed that since the laws of physics are 
invariant under time
reversal, it might be possible under certain circumstances to impose the condition that there is 
particular density matrix $\rho_f$ in the future. In the case of gravity, where distinguishing between the 
past and future is not straightforward, the impetus to do so is greater. They suggested the modification
\begin{equation} P[\rho_f,\rho_i] =
 {\mathcal N}\ {\rm Tr}\ \int {\mathcal D}g^{(+)}\ {\mathcal D}g^{(-)}\ \rho^{(f)}e^{iI[g^{(+)]}}
 \rho^{(i)} e^{-iI[g^{(-)}]}. \end{equation}
where  the two path integrals are over spacetime metrics that have spatial
metrics reproducing the density matrices $\rho^{(i)})$ and $\rho^{(f)}$ in the past and future
respectively.
Setting $\rho^{(f)}$ to the identity reproduces the usual formulation of quantum mechanics and 
has been termed by Gell-Mann and Hartle \cite{GellMann:1991ck} as the  principle of indifference. 
Specifying a non-trivial $\rho_f$ is termed post-selection. 

Some time ago, 
Horowitz and Maldacena \cite{Horowitz:2003he}
suggested this
type of boundary condition in the future might be able to resolve the information paradox. 
In fact, it seems that in order to prevent information leaking out of the spacetime through the singularity,
setting a boundary condition there is a necessity. Setting boundary conditions in the future risks
unusual behavior such as apparent violations of no-cloning, unitarity and causality,
\cite{Gottesman:2003up}.

If we thought that the approach to singularity was smooth, as seems to be the case for 
Oppenheimer-Snyder collapse, one would be wrong. Belinsky, Khalatnikov and E.M.Lifshitz (BKL) 
\cite{Belinsky:1982pk} showed 
that collapse is generally chaotic. We illustrate this for the case of pure gravity but the 
extension to include other fields is straightforward.. 
Classically, as we approach a spacelike singularity, space 
starts to break up into individual regions that do not interact with each other. 
Each of these regions behaves a bit like a Kasner spacetime. The Kasner spacetime has metric
\begin{equation} ds^2 = -dt^2 + a^2(t)dx^2 + b^2(t)dy^2 + c^2(t)dz^2 \end{equation}
with $a(t)=(t_0-t)^{p_1}, b(t)=(t_0-t)^{p_2}, c(t)=(t_0-t)^{p_3}$ and $t\le t_0$. The singularity is 
reached at $t=t_0$. The exponents $p_i$ obey $p_1+p_2+p_3=p_1^2+p_2^2+p_3^2=1$ so that one of the
$p_i$ is negative and two are positive. Two dimensions of space are contracting as one moves 
towards the singularity and one is expanding. Any object approaching the singularity will therefore
undergo spaghettification. Note that the volume of space is decreasing as one gets close to the 
singularity. BKL showed that a region undergoes Kasner behavior for a period but is then interrupted by 
two different types of process. The first is that the Kasner exponents are not fixed but experience a
rapid change at certain intervals. The second is that the principal directions of expansion and contraction
$x,y$ and $z$ undergo  rotation from time to time. The complicated behavior found by BKL can be 
succinctly summarised in a way first found by Damour, Henneaux and Nicolai, \cite{Damour:2002et}.
The Einstein equations amount to null geodesic motion in a space with metric
\begin{equation} 
d\sigma^2 = -d\rho^2 + \rho^2\ \Bigl(\frac{du^2+dv^2}{v^2}\Bigr). \label{eq:line}
\end{equation}
The $z=u+iv$ plane is a space of constant negative curvature. Motion is however restricted to 
the domain $F$ in the $z$ plane bounded by the unit circle centered on the origin and the lines $u=0, v>1$
and $u=1/2, v>\sqrt{3}/2$. This is precisely half of the more familiar fundamental region for
$SL(2,{\bf Z})$. When a null geodesic meets the boundary of $F$, it bounces
off it by specular reflection. Approach to the singularity corresponds to  $\rho\rightarrow\infty$.
Each point in this space corres[ponds to some three-geometry and the null geodesic sweeps out a curve
that represents the time evolution of this geometry.
In this picture, the Hamiltonian constraint takes the form
\begin{equation} {\mathcal H} = \frac{1}{2}\Biggl(-\pi_\rho^2 + 
\frac{v^2}{\rho^2}\ (\pi_u^2+\pi_v^2)\Biggr).
\end{equation}
Here $\pi_\rho,\pi_u$ and $\pi_v$ are the momenta conjugate to $\rho,u$ and $v$. 
Treating ${\mathcal H}$ as the Hamiltonian, together with the constraint ${\mathcal H}=0$ and the reflective
conditions 
at the boundary of $F$, reproduces the solution to the Einstein equations close to the singularity.

This system is quantized by replacing $\pi_\rho,\pi_u$ and $\pi_v$ by $-i{\partial/\partial \rho},
-i{\partial/\partial u}$ and $-i{\partial/\partial v}$ respectively. The Hamiltonian constraint then turns 
into the Wheeler-DeWitt equation on a minisuperspace. There is an operator ordering ambiguity 
in carrying this out,
so we have adopted an ordering such that the Wheeler-DeWitt equation is the Laplacian for the
metric in (\ref{eq:line}),
\begin{equation} \Bigl(-\rho^2\frac{\partial^2}{\partial \rho^2} - 2\rho\frac{\partial}{\partial\rho} -\Delta_F\Bigr)
\Psi = 0 \label{eq:wdw} \end{equation}
$\Delta_F$ is the Laplacian on the domain $F$ and is given by
\begin{equation} \Delta_F = -v^2 \Bigl(\frac{\partial^2}{\partial u^2} + 
\frac{\partial^2}{\partial v^2}\Bigr). \end{equation}
Since the walls are classically reflective, the 
boundary condition on $\Psi$ is that it vanishes on the boundary of $F$. 
The eigenfunctions $f_n$ of $\Delta_F$  obey
\begin{equation} \Delta_F f_n = s_n(1-s_n)f_n. \label{eq:eigen} \end{equation}
The $f_n$ are the odd Maass waveforms of $SL(2,{\bf Z})$ with $s_n = \tfrac{1}{2}\pm it_n, t_n$ real
and come in complex conjugate pairs. 
The spectrum of $\Delta_F$ has two distinct components; a discrete 
set of eigenfunctions that are the odd cusp forms of $SL(2,{\bf Z})$ and a continuum of the odd 
non-holomorphic Eisenstein series (NHES), \cite{Terras:xx}. The cusp forms are square integrable in $F$
whereas the NHES are not. Despite this, square integrable solutions of   
(\ref{eq:eigen}) can be written as a linear combination of the $f_n$, \cite{Terras:xx}. Each $f_n$ 
provides a solution of ({\ref{eq:wdw}), $\psi_n=\rho^{-s_n}f_n$. Hence a general solution of (\ref{eq:wdw})
is a linear combination $\psi_n$.  
The wavefunctions are not square integrable in  
minisuperspace even if they are square integrable in $F$. Such wavefunctions have appeared
in the literature before in a description of the initial cosmological singularity \cite{Kleinschmidt:2009cv}.
The natural inner product on 
functions in superspace is an analog of the Klein-Gordon norm \cite{DeWitt:1967yk}.
On minisuperspace, the inner product of two wavefunctions is
\begin{equation} (\Psi_a,\Psi_b) =i  \int_{\Sigma} \bigl(\Psi_a^\star\frac{\partial\Psi_b}
{\partial\rho}
 -\Psi_b^\star\frac{\partial\Psi_a}{\partial\rho}\bigr) \ \frac{\rho^2}{v^2}\ du\, dv \end{equation}
with $\Sigma$ being a \lq\lq spacelike\rq\rq\ surface of $\rho=$constant in the minisuperspace. 
By virtue of the Wheeler-DeWitt equation,
this is independent of $\rho$. Using this norm, any wavefunction that is integrable in $F$ descends to 
one with finite norm.

 As one
approaches the singularity where $\rho\rightarrow\infty$, the wavefunction behaves like $\rho^{-1/2}$.
The wavefunction therefore vanishes at the singularity,
a condition that was proposed by DeWitt,
 \cite{DeWitt:1967yk} as necessary for singularities to make sense quantum mechanically. 
 $(\Psi,\Psi)$ is in some sense the probability flux for gravitational information flowing through
 $\Sigma$. FOr generic $\Psi_a$, $(\Psi_a,\Psi_b)$ does not vanish as $\rho\rightarrow\infty$ even 
 though $\Psi \sim \rho^{-1/2}$.  However, one can choose perfectly reflecting boundary conditions 
 at the singularity,
 the information falling into the singularity is perfectly reflected. Under these circumstances
 $\Psi$ is real and $(\Psi,\Psi)=0$. Again this possibility was anticipated by DeWitt 
 as a mechanism for singularity avoidance \cite{DeWitt:1967yk}. This is 
 a boundary condition that specifies $\rho_f=0$. Under these circumstances, information is 
 not lost from the 
 spacetime and opens the possibility of rescuing unitary time evolution for black holes. 
 As spatial geometries approach the singularity one concludes that the probability of finding them is 
 going to zero because of the perfectly reflecting boundary conditions there.
 The approximation used will be accurate only for geometries close to being singular because 
 it is only there that the BKL walls become exactly reflective \cite{Belinsky:1982pk,Damour:2002et}.

We have seen how to set a future boundary condition
for the singularity, namely that $\rho_f=0$. 
 Outside the black hole, we assume that the principle of 
indifference holds. Such behavior gives a way of 
resolving the information paradox.

The singularity is required to reflect anything incident on it. One might be able to think of particles
getting close to the singularity as being annihilated by their antiparticles which travel backwards in time
and once they get outside the horizon scatter again and become outgoing Hawking radiation. In
some sense what happens close to the singularity is the time reverse of the Hawking pair creation process
in which particle-antiparticle pairs are created outside the horizon. 
The interior of the black hole is therefore a strange place where ones classical notions of
causality and unitarity are violated. This does not matter as long as outside the black hole
such pathologies do not bother us.
Lloyd \cite{Lloyd:2004wn} has found that classical information completely escapes 
from the black hole and
that quantum information mostly does. He estimated that at most half a qubit of information would 
be lost as the black hole disappears. Subsequently Lloyd and Preskill \cite{Lloyd:2013bza} showed that 
it was likely that both causality violation and unitarity could occur outside the black hole, 
but in practice it was most probably  
unobservable. We therefore conjecture that there is a quantum cosmic censor that forbids 
strange behavior outside the black hole. A slightly weaker form of this conjecture would be one 
that forbids the {\it observation} of strange behavior outside the black hole. 
One should also note that the conditions under which the firewall paradox was derived, do not hold 
for systems with post-selection, \cite{Almheiri:2012rt}. We should also note an intriguing connection
with number theory based on the nature of the wavefunctions in $F$.

\begin{acknowledgments}
I would like to thank the UK STFC for financial support under grant ST/L000415/1. 
I have enjoyed stimulating discussions with David Berman, Sam Braunstein, Jeremy Butterfield, 
Mihalis Dafermos, David Garfinkle, Gary Gibbons, Hadi Godazgar, Mahdi Godazgar, David Gross,
Stephen Hawking, Frans Pretorius, 
Maria Rodriguez, Edward Witten 
and Anna \.Zytkow
\end{acknowledgments}

\vfill\eject
\bibliography{info.bib}

\end{document}